\documentclass{llncs}
\usepackage{makeidx}  

\usepackage[dvips]{xcolor}
\usepackage{amssymb}
\usepackage{amsmath}
\begin{document}
\frontmatter          
\pagestyle{headings}  

\mainmatter

\title{Physical aspects of oracles for randomness, and Hadamard's conjecture}
\titlerunning{Oracles for randomness}  
%
\author{Karl Svozil}
\authorrunning{Karl Svozil} 
%
\tocauthor{Karl Svozil}
\institute{Institute for Theoretical Physics, Vienna University of Technology, \\
    Wiedner Hauptstra\ss e 8-10/136, A-1040 Vienna, Austria  \\
\email{svozil@tuwien.ac.at},
url:
\texttt{http://tph.tuwien.ac.at/$\sim$svozil}
}
\maketitle              

\begin{abstract}
We analyze the physical aspects and origins of currently proposed oracles for (absolute) randomness.
\end{abstract}

\section{Metamathematical and metaphysical origin of oracles for randomness}

Jozef Gruska's extensive reviews of the {\em foundations of computing}~\cite{Gruska-foc},
and {\em quantum computing}~\cite{Gruska}
documents his continued interest in the foundations of, and the connections between,
computation and physics.
This encouraged me to contribute to the physics of computation, in particular, by discussing
non-algorithmic oracles for randomness certified by physical principles.

The very existence of physical unknowables~\cite{svozil-07-physical_unknowables}
and indeterminism is subject to an ongoing debate that can be expected not to terminate at any time soon.
Thereby, like {\it Odysseus} trapped between {\it Scylla} and {\it Charybdis},
our perception of how the universe is organized has been vacuously oscillating between, and irritated by, claims of
physical determinism on the one hand, as well as complete indeterminism on the other hand.

Rather than arguing for one side or another, I would like to state upfront that both positions are metaphysical;
more precisely: from a physical perspective, these claims are non-operational.
And, formally, by reduction to the {\em halting problem}~\cite[Sec.~642]{Gruska-foc},
both of them are provable unprovable.
Because, form a purely phenomenological point of view, that is, in terms of the symbolic behaviour of physical systems,
any proof of determinism would imply solvability of the {\em rule inference problem}~\cite{angluin:83},
as well as total predictability
even beyond the {\em Busy Beaver} bound~\cite{chaitin-bb}.
Likewise, any claim of total indeterminism encounters the problem of enumerating an infinity of ``candidate theories of everything''~\cite{barrow-TOE},
let alone their future behaviour, as mentioned earlier.

Nevertheless, one way of corroborating physical indeterminism,
which could then be used for the construction of evidence-based oracles for randomness,
would be to ``screw open'' physical boxes which allegedly produce random bits.
We may not be able to do so, because, say,
relative to certain physical assumptions and formal theorems such as complementarity and value indefiniteness,
``nothing could be in'' such boxes~\cite{specker-60,2013-KstLip}.
But even then we may, at least, put forward
some theoretical arguments which are based on what we are inclined to believe~\cite[866]{born-26-1}.
In what follows we shall do exactly this:
we mention such oracles for randomness; that is, some boxes containing allegedly
indeterministic physical resources,
and why we believe (or not believe) that they act as physical sources of random bits.

A necessary and sufficient condition for this is the
existence of {\em gaps in the natural laws,} as discussed by Frank~\cite[Chapter~{III}, Sec.~12]{frank,franke}.
Such gaps allow, or rather necessitate, ``unlawful behaviour'' which could be utilized for physical oracles of randomness.

\section{Spontaneous symmetry breakdown and deterministic chaos}

Already in 1873, Maxwell identified a certain kind of {\em instability} at {\em singular points}
as rendering a gap in the natural laws \cite[211-212]{Campbell-1882}:
{\em ``$\ldots$~when an infinitely small variation in the present state may bring about a finite difference in the state of the
system in a finite time, the condition of the system is said to be unstable.
It is manifest that the existence of unstable conditions renders impossible the prediction of future events, if our
knowledge of the present state is only approximate, and not accurate.
$\ldots$~the system has a quantity of potential energy, which is
capable of being transformed into motion, but which cannot begin to be so transformed till the system has reached
a certain configuration, to attain which requires an expenditure of work, which in certain cases may be
infinitesimally small, and in general bears no definite proportion to the energy developed in consequence thereof.''}

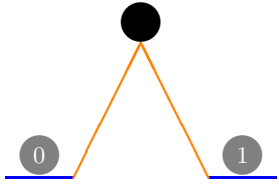
\begin{figure}
                \begin{centering}
\unitlength 3mm 
\linethickness{0.4pt}
\ifx\plotpoint\undefined\newsavebox{\plotpoint}\fi 
\begin{picture}(9,7)(0,0)
\thicklines
\put(0,0){\color{blue}\line(1,0){3.0}}
\put(9,0){\color{blue}\line(1,0){3.0}}
\put(3,0){\color{orange}\line(1,2){3.0}}
\put(9,0){\color{orange}\line(-1,2){3.0}}
\put(6,6.9){\color{black}\circle*{2}}
\put(1.5,1){\color{gray}\circle*{2}}
\put(10.5,1){\color{gray}\circle*{2}}
\put(1.5,1){\color{white}\makebox(0,0)[cc]{$0$}}
\put(10.5,1){\color{white}\makebox(0,0)[cc]{$1$}}
\end{picture}
                \end{centering}
                \caption{(Color online) A gap created by a black particle sitting on top of a potential well.
The two final states are indicated by grey circles. Their positions can be coded by $0$ and $1$, respectively.}
                \label{fig:2014-fw-instability}
        \end{figure}

Fig.~\ref{fig:2014-fw-instability} depicts a one dimensional gap configuration envisioned by Maxwell: a
{\em ``rock loosed by frost and balanced on a singular point of the mountain-side, the little spark which
kindles the great forest,~$\ldots$''}
On top, the rock is in perfect balanced symmetry.
A small perturbation or (pressure or thermal) fluctuation causes this symmetry to be broken,
thereby pushing the rock either to the left or to the right hand side of the potential divide.
This dichotomic alternative can be coded by $0$ and by $1$, respectively.

One may object to this scenario of {\em spontaneous symmetry breaking}
by maintaining that, if indeed the symmetry is perfect, there is no movement,
and the particle or rock stays on top of the tip (potential).
Any slightest movement might either result from a microscopic asymmetry of the initial state of the particle,
or from fluctuations of any form, either in the particle's position,
or by the surrounding environment of the particle.
For instance, any collision of gas molecules with the rock may push the latter over the edge
by thermal fluctuations.
Therefore, the randomness resides in the fluctuations, amplified by the instability.
Whether or not any such fluctuation may be considered as creating a gap is a question related to debates in statistical physics mentioned later.

A somewhat related scenario is that of {\em deterministic chaos,} because,
as Poincar{\'e} pointed out \cite[Chapter~4, Section~2,  p.+56--57]{poincare14}
{\em ``it can be the case that small differences in the initial values
produce great differences in the later phenomena;
a small error in the former may result in a large error in the latter.
The prediction becomes impossible and we have a ``random phenomenon.''}

\section{Quantum beam splitter}

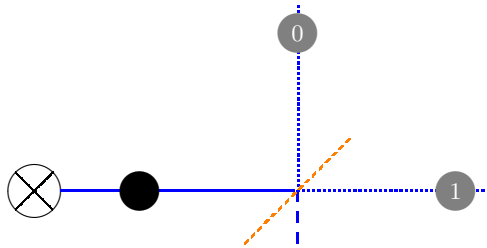
\begin{figure}
                \begin{centering}
\unitlength 0.7mm 
\linethickness{0.4pt}
\ifx\plotpoint\undefined\newsavebox{\plotpoint}\fi 
\begin{picture}(94.75,44.875)(0,0)

\put(10,10){\circle{10}}
\thinlines
\multiput(6.452,6.452)(.033885714,.033671429){210}{\line(1,0){.033885714}}
\multiput(13.523,6.452)(-.033880952,.033671429){210}{\line(-1,0){.033880952}}

\thicklines
{\color{blue}
\put(60,10){\line(0,-1){2}}
\put(60,6){\line(0,-1){2}}
\put(60,2){\line(0,-1){2}}
\put(15,10){\line(1,0){44.5}}
\multiput(60,9.93)(.972222,0){37}{{\rule{.8pt}{.8pt}}}
\multiput(60,9.805)(0,.972222){37}{{\rule{.8pt}{.8pt}}}
}
\put(30,10){\color{black}\circle*{8}}
\put(60,40){\color{gray}\circle*{8}}
\put(90,10){\color{gray}\circle*{8}}
\put(60,40){\color{white}\makebox(0,0)[cc]{$0$}}
\put(90,10){\color{white}\makebox(0,0)[cc]{$1$}}

\thicklines
{\color{orange}
\multiput(49.93,-.07)(.0322581,.0322581){20}{\line(1,0){.0322581}}
\multiput(51.22,1.22)(.0322581,.0322581){20}{\line(1,0){.0322581}}
\multiput(52.51,2.51)(.0322581,.0322581){20}{\line(1,0){.0322581}}
\multiput(53.801,3.801)(.0322581,.0322581){20}{\line(0,1){.0322581}}
\multiput(55.091,5.091)(.0322581,.0322581){20}{\line(0,1){.0322581}}
\multiput(56.381,6.381)(.0322581,.0322581){20}{\line(1,0){.0322581}}
\multiput(57.672,7.672)(.0322581,.0322581){20}{\line(1,0){.0322581}}
\multiput(58.962,8.962)(.0322581,.0322581){20}{\line(0,1){.0322581}}
\multiput(60.252,10.252)(.0322581,.0322581){20}{\line(0,1){.0322581}}
\multiput(61.543,11.543)(.0322581,.0322581){20}{\line(0,1){.0322581}}
\multiput(62.833,12.833)(.0322581,.0322581){20}{\line(1,0){.0322581}}
\multiput(64.123,14.123)(.0322581,.0322581){20}{\line(1,0){.0322581}}
\multiput(65.414,15.414)(.0322581,.0322581){20}{\line(0,1){.0322581}}
\multiput(66.704,16.704)(.0322581,.0322581){20}{\line(0,1){.0322581}}
\multiput(67.994,17.994)(.0322581,.0322581){20}{\line(0,1){.0322581}}
\multiput(69.285,19.285)(.0322581,.0322581){20}{\line(0,1){.0322581}}
}
\end{picture}
                \end{centering}
                \caption{(Color online) A gap created by a quantum coin toss. A single quantum (symbolized by a black circle
from a source (left crossed circle)
impinges on a semi-transparent mirror (dashed line), where it is reflected and transmitted with a 50:50 chance.
The two final states are indicated by grey circles. The exit ports of the mirror can be coded by $0$ and $1$, respectively.}
                \label{fig:2014-fw-qcointoss}
        \end{figure}

A quantum mechanical gap can be realized by a beam splitter~\cite{green-horn-zei,Weihs-2001}, such as a {\em half-silvered mirror},
with a 50:50 chance of transmission and reflection~\cite{svozil-qct,stefanov-2000,zeilinger:qct},
as depicted in Fig.~\ref{fig:2014-fw-qcointoss}.
A gap certified by quantum value indefiniteness necessarily has to operate with more than two exclusive outcomes~\cite{PhysRevA.89.032109}.
Ref.~\cite{2012-incomput-proofsCJ} presents such a qutrit configuration.

One may object to this scenario of {\em quantum indeterminism} by pointing out
that it is merely based on a believe
--
actually, Born's {\em inclinations
``to give up determinism in the world of atoms''}~\cite[p.~866]{born-26-1}
(English translation in \cite[p.~54]{wheeler-Zurek:83})
--
with provable formal improvability~\cite{svozil-2013-omelette}.
We shall come back to related issues later.

One may also object that a lossless beam splitter has a quantum mechanical
representation as an invertible unitary operator~\cite{zeilinger:882,green-horn-zei,rzbb} $\textsf{\textbf{U}}$,
and therefore is reversible.
Indeed, this can be readily demonstrated operationally by serially composing a lossless Mach-Zehnder interferometer with two beam splitters,
thereby reconstructing the original quantum state (signal); that is, more formally,
$\textsf{\textbf{U}}^\dagger \textsf{\textbf{U}}=\textsf{\textbf{I}}$, where ``$\dagger$''
indicates the Hermitian adjoint, and $\textsf{\textbf{I}}$ stands for the identity operator.
How this kind of unitarity conforms with the view that a beam splitter can be considered an ``active element'' of quantum randomness remains unresolved,
and is actually highly questionable~\cite{everett,wigner:mb}.
Often vacuum fluctuations originating from the second, empty, input port are mentioned,
but, pointedly stated~\cite[p.~249]{chau}, these {\em ``mysterious vacuum fluctuations $\ldots$ may be regarded as sugar coating for the bitter pill of
quantum theory.''}

A lossless 50:50 beam splitter can be modelled by a normalized $2 \times 2$
Hadamard transformation $\textsf{\textbf{U}}=  \frac{1}{\sqrt{2}}  \textsf{\textbf{H}}_2$
with rows $( \frac{1}{\sqrt{2}} , \frac{1}{\sqrt{2}} )$ and  $(- \frac{1}{\sqrt{2}} , \frac{1}{\sqrt{2}} )$, respectively.

More generally,
suppose we would like to construct a
$\underbrace{\frac{1}{n} : \frac{1}{n} : \ldots : \frac{1}{n}}_{n \text{ times}}$
beam splitter represented by a normalized  Hadamard matrix $\frac{1}{\sqrt{n}} \textsf{\textbf{H}}_n$;
that is, an Hadamard matrix $\textsf{\textbf{H}}_n$ divided by the square of (the dimension) $n$.
An $n \times n$ Hadamard matrix $\textsf{\textbf{H}}_n$ has entries in $\{-1,1\}$ such that any two distinct rows or columns of
$\textsf{\textbf{H}}_n$, interpreted as vectors in a Hilbert space,
have scalar product zero; that is, they are orthogonal
(or, equivalently, by requiring that its transpose $\textsf{\textbf{H}}_n^T$ satisfies $\textsf{\textbf{H}}_n \textsf{\textbf{H}}_n^T= n\textsf{\textbf{I}}_n$).

A {\em necessary} condition for such a construction~\cite{tressler-hadamard,horadam} is that $n=1$, $n=2$, or $n=4k$ for any $k\in {\Bbb N}$.
{\em Hadamard's conjecture} claims that this is also a {\em sufficient} condition for the existence of an $n$-dimensional Hadamard transformation;
and thus, for a corresponding equi-decomposition of quantum states into coherent superpositions.
(Of course, a quantum state can be decomposed into any fraction of unity by suitable unitary transformations; this just represents a permutation of the original state,
or, in a different interpretation, a base change~\cite{Schwinger.60}.)

A quantum oracle to Hadamar's conjecture would be one which would, for any $k \in {\Bbb N}$, output $4k$ orthogonal
$\frac{1}{4k}$-equi-weighted mixtures of orthogonal states spanning the entire
$4k$-dimensional (real) Hilbert space.
A beam splitter realizing Hadamard's conjecture would possess the remarkable property that it converts a signal input in any one of the
$4k$ input ports into a coherent equi-superposition of all output ports;
with relative phase differences equal to $0$ (corresponding to equal relative sign),
and $\pi$ (corresponding to  relative sign ``$-$'').

At the same time,
in terms of quantum states forming bases (or, by other namings, blocks, subalgebras or contexts~\cite{svozil-2008-ql}),
Hadamard's conjecture translates into the existence of a particular kind of pure states
equivalent to the projectors corresponding to the row (column) vector of a normalized Hadamard matrix.
The set of row vectors of $\frac{1}{\sqrt{4k}}\textsf{\textbf{H}}_{4k}$
correspond to an orthogonal basis which is {\em (mutually) unbiased} with respect to the
Cartesian standard basis in ${\Bbb R}^{4k}$.

Schwinger's construction~\cite{Schwinger.60} can be used for the rendition of mutually unbiased bases
in arbitrary dimensions $n$; alas the base vectors may have complex coordinates.
The construction starts with the Cartesian standard basis
$\{ \vert e_1 \rangle,  \vert e_2 \rangle,\ldots , \vert e_n \rangle \}$ and involves three steps:
(i) a cyclic shift of the basis vectors $\{ \vert f_1= e_2 \rangle,\ldots , \vert f_{n-1}=e_n \rangle , \vert f_n=e_1 \rangle  \}$,
(ii) the construction of a unitary operator
$\textsf{\textbf{U}}$ by $\textsf{\textbf{U}}=\sum_{i=1}^n  \vert e_i \rangle  \langle f_i \vert$; and finally
(iii) the identification of the normalized eigenvectors of $\textsf{\textbf{U}}$ with
the elements of a basis which is unbiased with respect to the Cartesian standard basis.
The  associated normalized complex Hadamard matrix~\cite{Bengtsson-07,PhysRevA.79.052316,Dita-2010}
is just the row (column) matrix of the elements of this basis.
For the sake of an example, we can readily write an algorithm~\cite{2012-schwinger.m} yielding a
complex Hadamard matrix of dimension 8; that is,
$$
\left(
\begin{array}{cccccccc}
 1 & 1 & 1 & 1 & 1 & 1 & 1 & 1 \\
 -1 & 1 & -1 & 1 & -1 & 1 & -1 & 1 \\
 i & -1 & -i & 1 & i & -1 & -i & 1 \\
 -i & -1 & i & 1 & -i & -1 & i & 1 \\
 (-1)^{1/4} & i & (-1)^{3/4} & -1 & -(-1)^{1/4} & -i & -(-1)^{3/4} & 1   \\
 -(-1)^{3/4} & -i & -(-1)^{1/4} & -1 & (-1)^{3/4} & i & (-1)^{1/4} & 1   \\
 (-1)^{3/4} & -i & (-1)^{1/4} & -1 & -(-1)^{3/4} & i & -(-1)^{1/4} & 1   \\
 -(-1)^{1/4} & i & -(-1)^{3/4} & -1 & (-1)^{1/4} & -i & (-1)^{3/4} & 1   \\
\end{array}
\right)
.
$$
Whether the Schwinger construction, for $n=4k$, $k\in {\Bbb N}$,
can be extended to produce only the real entries in $\{-1,1\}$
instead of complex numbers of modulus unity remains unknown.
One may conjecture that in this case the Dita decomposition~\cite{Dita-2003}
of unitary matrices into products of diagonal phase matrices (with modulus one entries)
and orthogonal matrices -- which in turn can be written as compositions of
rotations  in two-dimensional subspaces --
yields the appropriate real Hadamard matrices
by substituting $0$ or $\pi$ for all for all phases in the phase matrices (thereby rendering
diagonal elements
$1$ and $-1$, respectively), as well as by
identifying all rotation angles with $\pm \pi/4$ (thereby rendering factors whose absolute value is
$1/\sqrt{2}$).

\section{Quantum vacuum fluctuations}

As stated by Milonni~\cite[p.~xiii]{milonni-book} and others~\cite{einstein-aether,dirac-aether}, {\em ``$\ldots$~there is no vacuum in the ordinary sense of
tranquil nothingness. There is instead a fluctuating quantum vacuum.''}
One of the observable vacuum effects is the {\em spontaneous emission of radiation}~\cite{Weinberg-search}:
{\em ``$\ldots$~the process of spontaneous emission of radiation is one in which ``particles'' are actually created.
Before the event, it consists of an excited atom, whereas after the event, it consists of an atom in a state of lower energy, plus a photon.''}
Recent experiment achieve single photon production by spontaneous emission~\cite{PhysRevLett.39.691,PhysRevLett.85.290,Buckley-12,Stevenson-spontemi,Sanguinetti},
for instance by electroluminescence.
Indeed, most of the visible light emitted by the sun or other sources of blackbody radiation, including incandescent bulbs,
is due to spontaneous emission~\cite[p.~78]{milonni-book} and thus subject to {\em creatio ex nihilo}.

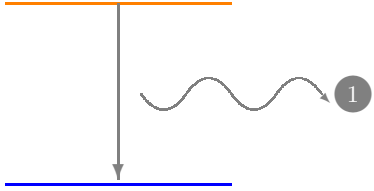
\begin{figure}
                \begin{centering}
\unitlength 0.6mm 
\linethickness{0.4pt}
\ifx\plotpoint\undefined\newsavebox{\plotpoint}\fi 
\begin{picture}(79.526,40)(0,0)
\thicklines
\put(0,0){\color{blue}\line(1,0){50}}
\put(0,40){\color{orange}\line(1,0){50}}
\put(25,40){\color{gray}\vector(0,-1){39}}
\thinlines
{\color{gray}
\qbezier(30,20)(35,13)(40,20)
\qbezier(70,20)(65,27)(60,20)
\qbezier(50,20)(45,27)(40,20)
\qbezier(50,20)(55,13)(60,20)
\put(70,20){\vector(1,-1){2}}
}
\put(77,20){\color{gray}\circle*{8}}
\put(77,20){\color{white}\makebox(0,0)[cc]{$1$}}
\end{picture}
                \end{centering}
                \caption{(Color online) A gap created by the spontaneous creation of a photon.}
                \label{fig:2014-fw-vacuumfluctuation}
        \end{figure}

A gap based on vacuum fluctuations is schematically depicted in Fig.~\ref{fig:2014-fw-vacuumfluctuation}.
It consists of an atom in an excited state, which transits into a state of lower energy, thereby producing a photon.
The photon (non-)creation can be coded by the symbols $0$ and $1$, respectively.


\section{Analogies in statistical physics}

In the following we shall briefly glance at two related physical issues
--
the purported (ir-)reversibility
of quantum measurements~\cite{PhysRevD.22.879,PhysRevA.25.2208,greenberger2,Nature351,Zajonc-91,PhysRevA.45.7729,PhysRevLett.73.1223,PhysRevLett.75.3783,hkwz},
as well the character of the second law of thermodynamics~\cite{Myrvold2011237}.

\subsection{Wigner's and Everett's arguments against quantum measurement}

The extension of the observation context is not dissimilar to what Wigner~\cite{wigner:mb}
and, in particular, Everett~\cite{everett,everett-collw} had in mind when they argued against (irreversible
and, in principal, for reversible) measurement.
Because quantum mechanics allows for two types of evolution:
(i) the first type comprises irreversible measurements,
whereas (ii) the second mode is characterized by the unitary, that is, reversible permutation, of quantum states
in-between aforementioned measurements.

Alas, this is true only
{\em for all practical purposes}~\cite{PhysRevD.22.879,PhysRevA.25.2208,greenberger2,bell-a,Nature351,Zajonc-91,PhysRevA.45.7729,PhysRevLett.73.1223,PhysRevLett.75.3783,hkwz},
that is, {relative to the physical means}~\cite{Myrvold2011237} available to resolve the huge number of degrees of freedom involving a
``macroscopic'' measurement apparatus.
And yet, at least in principle, if the unitary quantum evolution is taken to be universally valid,
then any distinction or cut between the observer and the measurement apparatus on the one side,
and the quantized object on the other side, is not absolute or ontic,
but epistemic, means-relative, subjective and conventional~\cite{svozil-2013-omelette}.

\subsection{Analogies to the second law of thermodynamics}

There are good reasons to believe that also irreversibility in statistical physics
is means relative~\cite{Myrvold2011237} and thus epistemic: if we cannot resolve individual constituents of a group, and their  degrees of freedom,
then irreversibility is the epistemic expression of our incapacity to do so.
In contradistinction, suppose the molecules are taken individually.
In this case the second law might ``dissolve into thin air'' because of reversibility on the micro-description level.
In Maxwell's own words~\cite[Document~15, p.~422]{garber}
{\em ``I carefully abstain from asking the molecules which enter where they last started
from. I only count them and register their mean velocities, avoiding all personal
enquiries which would only get me into trouble.''}

\section{{\it Caveats} and afterthoughts}

Stated pointedly,
we have essentially been talking about the emergence of events {\em out of nothing} (e.g. {\it creatio ex nihilo}) and
without any cause.
Thereby, and for the sake of accepting classical and quantum oracles for randomness,
we are denying the  {\em principle of sufficient reason,}
as well as negating Parmenides' {\em nothing comes from nothing,}
which so powerfully guided the ancient Greek and modern western Enlightenments.

More technically,
we note without further discussion that any ``diluted''~\cite{Calude200620} indeterminism, or gap mechanism,
could be ``concentrated'' to Borel normality
by assuming independence of bits in binary sequences~\cite{von-neumann1,AbbottCalude10}.

As a last  speculation,
it might not be too unreasonable to contemplate
that all gap scenarios,
including spontaneous symmetry breakdown and quantum oracles,
are ultimately based on vacuum fluctuations.

\section*{Acknowledgments}
This research has been partly supported by FP7-PEOPLE-2010-IRSES-269151-RANPHYS.


\end{document}